\documentclass{iaus}

\title[Graduate Program in Astrophysics in Split] 
{Graduate Program in Astrophysics in Split}

\author[Davor Krajnovi\'c]   
{Davor Krajnovi\'c$^1$%
  \thanks{Present address: University of Oxford, Keble Road, Oxford, OX1 3RH, UK},
}

\affiliation{$^1$Department of Physics, University of Oxford,
Oxford, UK\break email: dxk@astro.ox.ac.uk\\[\affilskip]}

\pubyear{2006}
\volume{SPS5}  
\pagerange{??--??}
\date{?? and in revised form ??}
\jname{Astronomy for the Developing world}
\editors{J. Hearnshaw \& P. Martinez}
\begin{document}

\maketitle

\begin{abstract}
Beginning in autumn 2008 the first generation of astronomy master
students will start a 2 year course in Astrophysics offered by the
Physics department of the University of Split, Croatia
(http://fizika.pmfst.hr/astro/english/index.html). This unique master
course in South-Eastern Europe, following the Bologna convention and
given by astronomers from international institutions, offers a series
of comprehensive lectures designed to greatly enhance students'
knowledge and skills in astrophysics, and prepare them for a
scientific career. An equally important aim of the course is to
recognise the areas in which astronomy and astrophysics can serve as a
national asset and to use them to prepare young people for real life
challenges, enabling graduates to enter the modern society as a
skilled and attractive work-force. In this contribution, I present an
example of a successful organisation of international astrophysics
studies in a developing country, which aims to become a leading
graduate program in astrophysics in the broader region. I will focus
on the benefits of the project showing why and in what way astronomy
can be interesting for third world countries, what are the benefits
for the individual students, nation and region, but also research,
science and the astronomical community in general.

\end{abstract}

\firstsection 
\section{Introduction}

The knowledge-based economy is founded on discoveries and
innovations. The power of a nation is measured by its ability to
stimulate discoveries and capacity to innovate. A nation cannot
achieve this without increasing its educational levels and
intellectual competences. Moreover, for a small nation, building
knowledge-based society is more than an economic necessity -- it is
also a strategy for preserving its culture and identity.

Astronomy, as a scientific discipline, is a valuable incubator of new
ideas, discoveries and enterprises, driven by the synergy of
sophisticated technologies, different natural sciences and the wish to
understand our Universe. Astronomy has also a unique public appeal,
which enables it to inspire young generations and stimulate their
curiosity, creativity and appreciation for science.

In developing countries, astronomy is, however, often perceived as an
abstract and expensive science, without visible end-products through
which it is possible to quantitatively 'weigh' its usefulness. Even
in developed countries, in the times of economic uncertainty,
astronomy falls into the group of fundamental sciences which are the
first to loose governmental monetary support.  It is the duty of
professional astronomers to oppose this opinion, and to demonstrate
the benefits related to astronomy as a science and, especially, as an
educational option for bright young people of all ages and school
levels.\looseness=-2

Astrophysics Initiative in Dalmatia (ApID) is an attempt to do this
and promote the growth and development of astrophysics in Dalmatia, a
southern region of Croatia. ApID is a collaboration of projects,
institutions, and individuals, from professional scientists to amateur
astronomers, sharing the common interest in astronomy and
astrophysics. The strategy of ApID is to ensure its sustainability
through a carefully designed combination of educational, research, and
outreach programs. The operational performance and strength of ApID
are enhanced by the synergy of a public university (the University of
Split) and a non-profit non-governmental organisation (the Society
znanost.org). The backbone of ApID is the Graduate Program in
Astrophysics at the Department of Physics at the University of Split
(GPAS), Croatia, which is the main topic of this contribution to the
Special Session 5. The main objectives of GPAS can be summarised as
follows:
\begin{itemize}

\item[-] provide the graduates with know-how to continue in diverse
  career paths as well as to make them attractive targets for a range
  of employers in different branches of business, government and
  finances,

\item[-] attract young people for careers in science and technology, 

\item[-] establish connections with other disciplines present at the
  University of Split, other Universities in Croatia and abroad in
  order to actively promote interdisciplinary sciences,

\item[-] promote top level research in astronomy in Croatia, and

\item[-] enhance the quality of existing graduate level education in the
  region and set an example to other educational institutions in South
  Eastern Europe.

\end{itemize}

Here, I outline the graduate program focusing on aspects which differ
from existing and more classical courses (Section~\ref{s:org}). This
is followed by a discussion about the benefits that such a study can
bring to the students, to the region, but also to us, the astronomical
community (Section~\ref{s:ben}). Finally, I conclude with highlights
of the GPAS (Section~\ref{s:con}).

\section{Study astrophysics in Split}
\label{s:org}

GPAS is supported by the Department of Physics of the University of
Split, which is a part of the Faculty of Natural Sciences, Mathematics
and Kinesiology. Recently, following the Bologna process
(standardisation of higher education in Europe), the Department of
Physics restructured its courses now offering three-year undergraduate
and two-year MSc (master’s degree) graduate study
programs. The MSc in Astrophysics will be given in English, enabling
students from other countries to follow the courses in Split
increasing their mobility and exchange of experiences.

{\bf Students.} The MSc program in Astrophysics is open to Croatian
and international students, who completed their undergraduates studies
(BSc) in physics or related subject. In the modern world, it is
necessary to actively work on attracting bright students. Clearly, the
best undergraduate students should be encouraged to apply to the
program. Possible candidates, however, could be identified and
attracted to astronomy at even younger age, among the gifted
high-school or even elementary schools pupils. This means that a part
in the organisation of the graduate studies also includes an active
outreach through which GPAS will be present in the media and news,
will actively participate in the organisation of local and regional
scientific events for young people, and will promote the
popularisation of science to the general public.

{\bf Lecturers.} Currently, there are no professional astronomers at
the Department of Physics in Split. The expanding department will in
the following years offer positions for astronomers, but the bulk of
the astronomy specific teaching at GPAS will be given by visiting
lectures. Some of these lecturers are Croatian astronomers working
abroad, while others are foreign scientists. The rest of the courses
(non-astronomy related, but necessary for a completion of the master
degree in Astrophysics) will be given by local physics faculty. The
vising lecturers will not need to spend a whole semester in Split, but
rather a few weeks necessary for completing their course.

{\bf Courses and Timetable}.  The courses are divided between required
and elective courses. The aim is to give a thorough astrophysical
background, which is supplemented with optional courses in general
physics, computer science and humanities (history, philosophy, etc).
Since large section of courses will be given by visiting lecturers,
the whole educational program is organised in blocks. One block,
lasting between 2 to 4 weeks, contains the lectures as well as the
exam. The emphasis of the studies on the research is evident in the
fact that in all semesters students will be required to participate in
scientific projects. Moreover, the last semester is completely devoted
to a research project which will become the master thesis. The aim of
these projects is to encourage students to tackle complex programs and
learn methods of the scientific research. The most important aspect,
however, is to enable students to work in an international environment
at astrophysical centres of excellence. They will have to work on
projects offered by the lecturers, visiting their institutions abroad
and producing publishable scientific results.

{\bf The first light.} The new graduate program will commence during
the fall of 2008 when the first group of Croatian students following
the Bologna process will finish their BSc. \looseness=-2

\section{Benefits of the graduate program}
\label{s:ben}

The position of astronomy in early societies was very much different
than it is today. First astronomical exploits were very practical,
from the determination of the seasons necessary for the agriculture,
which fed the stratified human societies, to predictions of terrifying
eclipses and explanation of the heavens. Astronomy was a very
practical and necessary system of theoretical knowledge, with deep
cultural and social consequences. Last few hundred years, however, saw
a reverse of the medal. Astronomy, supported by physical laws of
nature, was turning towards theoretical knowledge of Universe, its
structure and properties, loosing its practicality and usefulness for
everyday aspects of human lives. On the other hand, younger sciences
like physics, chemistry, biology and their various mixtures (medicine,
engineering...), with their technical applications, started changing
the human society as well as its influence on the Earth. Astronomy did
little in this 'progress'\footnote{One should, however, remember the
crucial role of astronomy in navigation and, hence, in the great
geographical discoveries and world trade until the recent advent of
satellite navigation. The importance of astronomy for economy of the
sea-faring nations is evident also in the support astronomical
research was getting from the governments: both Paris and Greenwich
Observatories were opened in 17$^{th}$ century with purpose to perfect
the art of navigation, increase the maritime power of the nations and,
in general, make the seafaring safer.}, becoming more and more
academic and idealistic pursuit of the nature of the Universe. Today,
it is often considered a fundamental science, remote and useless for
'real' life, perhaps an exercise for the mind, interesting in terms of
secular human culture; an entertainment, although rather expensive, of
general public.

Astronomy in the modern world, however, is much more than that. It
still makes humanistic, educational and technical contributions to our
society. Rather than going through a list of recent contributions from
astronomical research, I will focus on a more specific aspect of
benefits that a study of astronomy can offer to the students and to
the region hosting an astronomical institution.

\subsection{Benefits for the students}
\label{ss:stud}

It is self-evident that training in science opens doors to a
scientific career. It is, however, often not clear to the general
public that the same training can be very attractive to industry and
different businesses as potential employers of skilled people. An
objective of GPAS is to produce world-class graduates with the
necessary skills to become local and global leaders and entrepreneurs.

Industry and finances are generally not concerned about the
specialisation of an academic course. They are more interested in
broader skills obtained by students, which can be transferred to the
work place. A master course in astronomy is actually well suited in
this respect. The interdisciplinary nature of astronomy provides a
full framework for illustrating to the students the unity of natural
phenomena and the evolution of scientific paradigms that explain
them. On a more pedagogical level within higher education, teaching
astronomy to master students prepares them for a broad range of
scientific disciplines ranging from purely academic, such as
astrophysics, to widely applicable tasks of engineering and computer
science. GPAS aims to leave a number of transferable skills to its
graduates:

\begin{itemize}
\item[-] mathematical, computing and modelling skills with scientific literacy 
\item[-] tacking complex problems and handling incomplete and large data sets
\item[-] cooperating in international teams and projects 
\item[-] preparing and executing projects (often on tight deadlines) 
\item[-] presenting an account of work to colleagues and to broader audiences
\end{itemize}

\noindent It is important to stress that the above mentioned skills
provide a solid base not only for a career in science, but also in
different branches of business.  It is the synergy of different
sciences used in astronomical research that offers a broad spectrum of
transferable skills and know-how for diverse career paths.

\subsection{Benefits for the region}
\label{ss:reg}

The region concerned in this section is primarily Dalmatia with its
capital Split. On a slightly higher level it, of course, concerns
Croatia as the country where the master study is organised. It is,
however, easy to generalise towards a larger region of South--East
Europe, but certain aspects of the graduate program apply to the whole
Europe, both to its political and geographical entities.

Modern economy of a nation depends on its ability to compete
technologically with other nations. This ability directly translates
in the number of technically trained people that are able to use
existing technologies and develop new ones for international
markets. In addition to that, the quality of environment depends on
developing safe, clean industries and sources of energy. This, also,
can be accomplished only by imaginative and highly trained scientists
and engineers. However, even governments of developed countries
recognise the steadily decreasing number of students interested in
pursuing technical careers as a major concern for the future
development of their nations. To overcome this downward trend it will
be necessary to support those activities that stimulate young people
toward scientific thinking and the development of mathematical and
technical skills. The subject of astronomy is inherently interesting
to young people, thereby keeping them interested in science, whilst
they learn fundamentals of mathematics, statistics, physics,
chemistry, etc. GPAS follows the trends of physics departments in
developed counties which started offering astronomy options in their
curriculum, resulting in an increased interest in bachelor and master
degree programs in astronomy.  GPAS aims to provide a highly educated
and scientifically literate work force required by their home country
for its future development.

A big problem facing developing countries is the {\it brain-drain},
the exodus of highly educated individuals who go to wealthy nations
capable of offering well payed or other specific jobs (such as
research positions in astronomy). Croatia and the region of Split
suffer from this effect as well. A graduate course in astronomy,
however, can be used to reverse the {\it brain drain} into {\it brain
  gain} by attracting astronomy professionals (either Croatian or
foreign) to take posts in Split and promoting a modern program which
offers transferable skills to its students preparing them for
different career paths.

GPAS should also be interesting to the relevant bodies in the European
Union, since it is fully compatible with the European standards for
high education (Bologna process), emphasises the mobility of
students between the countries, especially the neighbouring countries
of Croatia, and its existence will enhance the quality of existing
graduate level education in the region, as well as set an example to
other educational institutions in South-Eastern Europe.

\subsection{Benefits for astronomy}
\label{ss:ast}

As a final remark in this section, it does not hurt to stress that with
a new graduate program in astronomy the whole astronomical community
also gains. An obvious benefits are the job openings and a
popularisation and spread of astronomy to the general public (who is
paying for most of the astronomical research). Possibly the most
interesting gain is opening of a new channel for introduction of young
bright people into the world of top level astronomical research. The
graduate program in Split supports this aspect by linking professional
scientists and students. This is achieved by bringing astronomers
(lectures, see Section~\ref{s:org}), who work at international
institutes and are involved in different international projects, to
Split to give lectures and offer master projects to the students. On
the other hand, they will meet capable students with different
backgrounds and eager to do astronomical research but coming from
Croatia and neighbouring countries which do not have developed
astronomical infrastructure nor are members of major international
projects or organisations. To our knowledge, the astrophysics graduate
course in Split is the first program of this scale offering the link
between astronomically developed western and underdeveloped
south-eastern Europe.

\section{Conclusions}
\label{s:con}

Astronomy is a lively natural science that can offer a broad range of
products to the general community: from 'pretty' pictures to an
understanding of the structure of the Universe, from technological
advancements to educational programs. Many of these are applicable and
useful for developing countries. In this work I focused on the
tertiary education, on the organisation of a graduate course in
astrophysics and on the benefits it offers to the students and the
host country. Here I summarise the main points. The aims of GPAS
are:\looseness=-2

\begin{itemize}

\item to recognise the areas in which astronomy and astrophysics can
  serve as a national asset and to use them to prepare young people
  for real life challenges;

\item to attract young people for careers in science and technology;

\item to set the example of excellence to be followed by other
  existing higher education programs in the region and demonstrating
  to the public that investing in science is the right thing to do;

\item to bring the top science and technology research to Croatia,
  enhance the transfer of technology to Croatia through international
  collaborations and joint projects with international centres of
  excellence, and help disseminate the skills and knowledge necessary
  for the establishment of similar world-class centres of excellence
  in the region; and 

\item to actively participates in the process of brain-gain by
  attracting Croatian science diaspora and foreign scientists to
  actively participate in the Croatian education system through the
  transfer of their skills and knowledge to Croatia.

\end{itemize}

The astrophysics graduate course is fully compatible with the European
educational standards, and its international orientation (curriculum,
language and lecturers) and research excellence will serve as a bridge
between the existing and future EU states, empowering Croatia as a
future EU member.

\begin{acknowledgments}
I would like to thank Dejan Vinkovi\'c, for useful discussions and
careful reading of the manuscript.
\end{acknowledgments}

\begin{discussion}

\end{discussion}


\begin{thebibliography}{}
\bibitem[]{}{} The following list of references have been consulted during the 
development of this document:\\
\end{thebibliography}
\end{document}